\documentclass[runningheads]{llncs}

\usepackage[width=122mm,left=12mm,paperwidth=146mm,height=193mm,top=12mm,paperheight=217mm]{geometry} 

\usepackage{graphicx}
\graphicspath{{figures/}}
\usepackage{amsmath,amssymb}
\usepackage{hyperref}
\DeclareMathOperator{\E}{\mathbb{E}}

\begin{document}

\title{Reinforcement Learning of Musculoskeletal Control from Functional Simulations}
\titlerunning{Reinforcement Learning for Biomechanical Simulations}

\author{Emanuel Joos \and Fabien P\'ean \and Orcun Goksel}
\authorrunning{E.\ Joos et al.}

\institute{Computer-assisted Applications in Medicine, ETH Zurich, Switzerland}

\maketitle

\begin{abstract}
To diagnose, plan, and treat musculoskeletal pathologies, understanding and reproducing muscle recruitment for complex movements is essential.  With muscle activations for movements often being highly redundant, nonlinear, and time dependent, machine learning can provide a solution for their modeling and control for anatomy-specific musculoskeletal simulations. Sophisticated biomechanical simulations often require specialized computational environments, being numerically complex and slow, hindering their integration with typical deep learning frameworks. In this work, a deep reinforcement learning (DRL) based inverse dynamics controller is trained to control muscle activations of a biomechanical model of the human shoulder.  In a generalizable end-to-end fashion, muscle activations are learned given current and desired position-velocity pairs.  A customized reward functions for trajectory control is introduced, enabling straightforward extension to additional muscles and higher degrees of freedom.  Using the biomechanical model, multiple episodes are simulated on a cluster simultaneously using the evolving neural models of the DRL being trained. Results are presented for a single-axis motion control of shoulder abduction for the task of following randomly generated angular trajectories.
\end{abstract}

\section{Introduction}
Biomechanical tissue models have been proposed for several different anatomical structures such as the prostate, brain, liver, and muscles; for various computer-assisted applications including preoperative planning, intraoperative navigation and visualization, implant optimization, and simulated training.
Musculoskeletal biomechanical simulations that use muscle activation models are used in orthopedics for functional understanding of complex joints and movements as well as for patient-specific surgical planning.
Shoulder is the most complex joint in the body, offering the greatest range-of-motion.
The upper arm is actively controlled and stabilized with over 10 anatomical muscles~\cite{Giacomo2008} subdivided in several parts~\cite{Brown2007}.
With high range-of-motion and redundancy, the shoulder is regularly exposed to forces larger than the body weight, making the area particularly prone to soft tissue damages~\cite{Streit2012,Craik2014}.
Consequent surgical interventions and corresponding planing and decisions could benefit from simulated functional models.

For simulating complex biomechanical models, sophisticated computational and simulation environments are often needed, such as SOFA~\cite{faure2012} and Artisynth~\cite{Lloyd}.
Due to many tortuous effects such as time-dependent, nonlinear behaviour of muscle fibres and soft tissue and the bone and muscle contacts and collisions, the control of muscle activations required for a desired movement is not trivial.
Linearized control schemes~\cite{Stavness2012} easily become unstable for complex motion and anatomy, e.g.\ the shoulder, despite tedious controller parametrization and small simulation time-steps leading to lengthy computations~\cite{Pean2019}.
Machine learning based solutions for such biomechanical models would not only enable simple, fast, stable, and thus effective controllers, but could also facilitate studying motor control paradigms such as neural adaptation and rehabilitation efficacy after orthopedic surgeries, e.g.\ muscle transfers.

Reinforcement learning (RL) is a machine learning technique for model control of complex behaviour, in a black-box manner from trial-and-error of input-output combinations, i.e.\ not requiring information on the underlying model nor its environment.
With Deep Reinforcement Learning (DRL), impressive examples have been demonstrated such as for playing Atari console games~\cite{Mnih2015} and the game of Go~\cite{Silver2016}, for control of industrial robots~\cite{james2016,Tsurumine2019}, and for animating characters~\cite{Lillicrap2015,Heess2017,kidzinski2018}, e.g.\ learning how to walk.
Despite DRL applications with simpler rigid and multi-body dynamics as above, soft-body structures and complex material, activation, geometry, and contact models have not been well studied.
Furthermore, sophisticated simulations required for complex models are not trivial to couple with DRL strategies.
In this paper, we present the DRL control of a Finite Element Method (FEM) based musculoskeletal shoulder model, while investigating two DRL approaches comparatively.

\section{Materials and Methods}
\begin{figure}[t]
\centering
\includegraphics[height=0.37\textwidth]{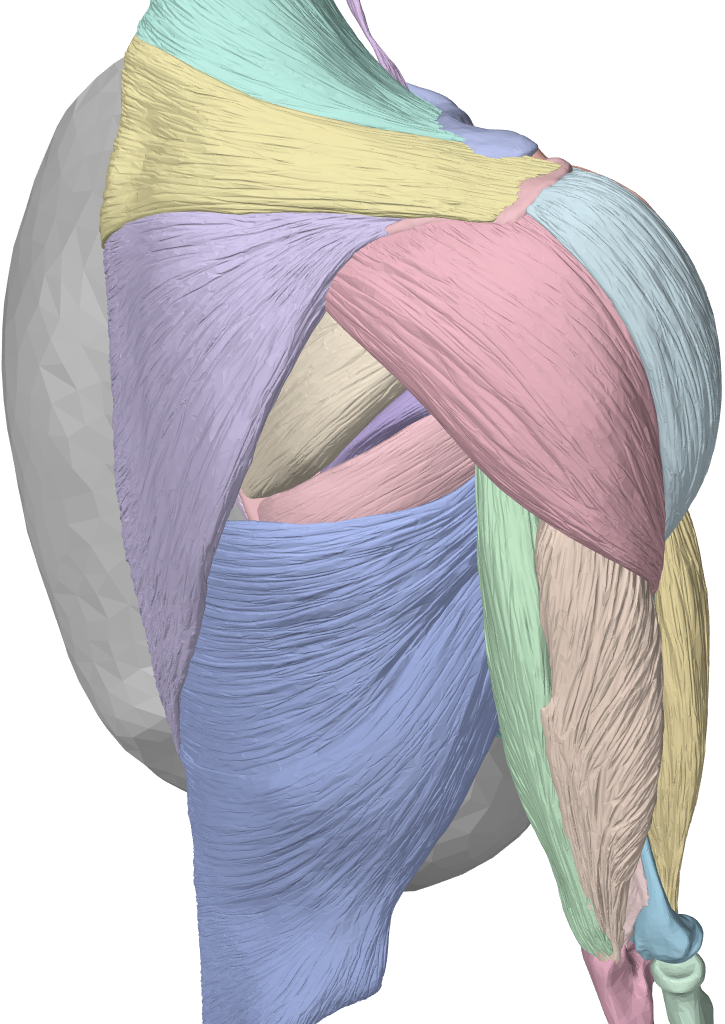}\hfill%
\includegraphics[height=0.37\textwidth]{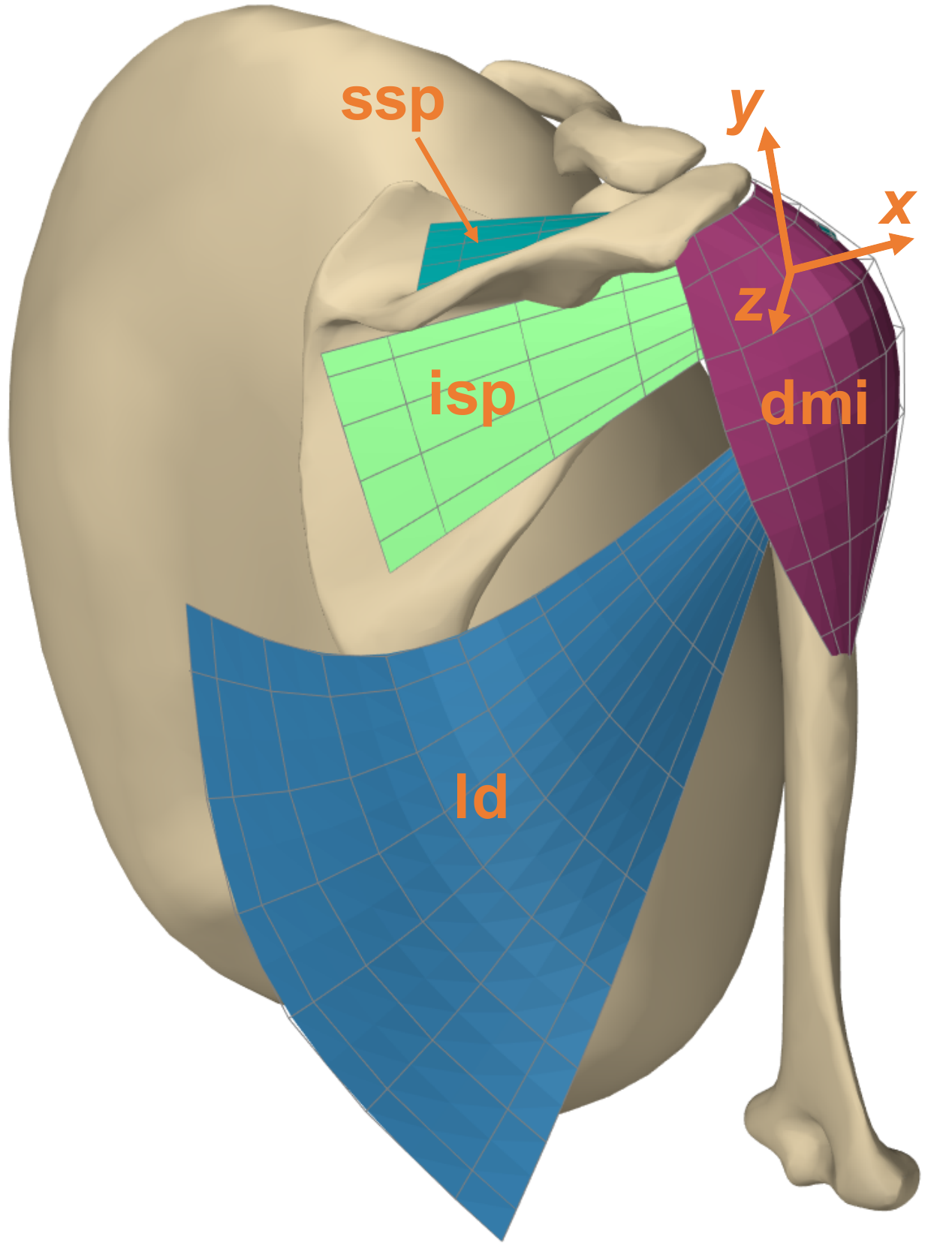}\hfill%
\includegraphics[height=0.35\textwidth]{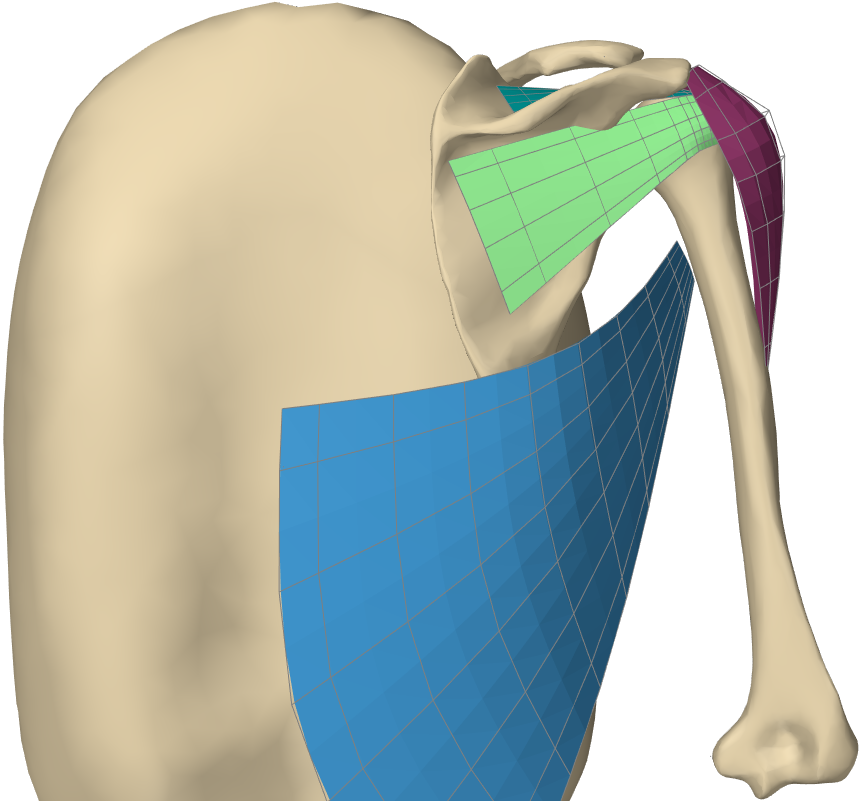}%
\caption{Shoulder segmentation~\cite{Mitsuhashi2008} (left) and our functional musculoskeletal model of the shoulder with four muscles relevant to abduction (center-right).} \label{MuskuloSkeletalModel}
\end{figure}

\noindent\textbf{Musculoskeletal Shoulder Model. }
We herein demonstrate DRL-based control with a simplified model of the shoulder joint.
We used segmentations from the BodyParts3D dataset~\cite{Mitsuhashi2008}, cf.\ Fig.\ref{MuskuloSkeletalModel}-left.
The shoulder complex consists of three bones: the humerus, the scapula and the clavicle; as well as multiple muscles, tendons and ligaments. 
Our model involves surface triangulations for the three bones; a manual surface fit to the ribs imitating the trunk; and B-spline based quadrilateral thin-shell meshes to model large, relatively flat muscles via FEM~\cite{Pean2020}.
The bones are rigid objects and the muscles are displacement-constrained on the bones at tendon origins and insertions.
Muscle fibres are modeled nonlinearly with respect to deformation based on~\cite{Blemker2005a}.
Fibres are embedded within a linear co-rotational FE background material model, coupled at FE integration nodes.
A normalized activation signal $a\in[0,1]$ sent homogeneously to all fibres of a muscle segment linearly generate internal stresses, contracting them against their background soft-tissue material, while pulling on the attached bones~\cite{Pean2020}.  
In this paper, we focus on the abduction motion, being a standard reference movement for diagnosis and in clinical studies~\cite{Wickham2010,Gerber2014,Contemori2019}.
Accordingly, four muscles relevant for abduction and adduction~\cite{Reed2013} supraspinatus (ssp), infraspinatus (isp), deltoid middle part (dmi), and latissimus dorsi (ld) are simulated herein, cf.\ Fig.\ref{MuskuloSkeletalModel}-center.

\vspace{1ex}\noindent\textbf{Learning Muscle Control. }
Consider that at each time step $t$, an agent with a current \emph{state} $s_t$ executes an \emph{action} $a_t$ according to some \emph{policy} $\pi(a_t|s_t)$, makes an observation $o_t$, and receives a scalar reward $r_t$.
RL aims to find the policy $\pi$ for an optimal outcome that maximizes the cumulative sum of current and discounted future rewards. 
This is predicted either based only on the current state, i.e.\ \emph{state value} function $V_{\pi}(s_t) = \E_{\pi}\left[r_t + \gamma r_{t+1} + \gamma^2r_{t+2} +... |s_t\right]$\,, or based on 
the current state and action together, i.e.\  \emph{action value} function  
$Q_{\pi}(s_t,a_t) = \E_{\pi}\left[r_t + \gamma r_{t+1} + \gamma^2r_{t+2} +... |s_t,a_t\right]$ also known as \emph{Q function}. 
$\gamma < 1$ is the discounting factor to ensure future rewards are worth lower. 

The shoulder model and its forward simulation is herein considered as black-box with its input being discrete muscles activation's and the output being the (angular) pose $\phi \in R^d$ and velocity $\phi'$ of the humerus, where $d$ is the degrees of freedom to control.
Using full 100\% activation range of the muscles as the potential RL action set makes it difficult to predict small activation changes precisely, as well as leading potentially to large activation jumps at failed predictions.
Therefore we use an action space of differential activation changes $\omega \in R^n$ where $n$ is the number of muscles controlled.
We thus additionally provide the current muscle activations $\Omega \in \mathbb{R}^n$ as input to the agent so that it can infer the incremental effect of $\omega$ on top.
To formalize a solution, typically a Markov Reward Process (MRP)~\cite{Bertsekas2012} is defined as a quintuple with the set of possible states $S$, the set of possible actions $A$, the reward function $r$, the transition probability matrix $P$, and a discounting factor $\gamma$.
Given the above, our state set is then 
\begin{equation}
    S \in \{ \phi(t), \phi'(t), \hat{\phi}(t+1), \hat{\phi}'(t+1), \vec{\Omega}(t)\},  
\end{equation}
where the hatted variables indicate the desired position and velocity at the next step.
Accordingly, we require only a short look-ahead, allowing for simpler network structures and efficient real-time inference.
We employ the following reward strategy
\begin{equation}
    \label{eq:reward}
     r(t+1) = -\left|\phi(t+1)- \hat{\phi}(t+1)\right|-\alpha \sum_{i=1}^{n}|\omega_i(t)| - \frac{1}{n}\sum_{i=1}^{n}
	1_{\{|\omega_i(t)|>\omega_\text{max}\}}.
\end{equation}
with the first term enforcing to follow the desired trajectory.
Lasso regularization in the second term encourages a sparse activation vector $\omega$, to resolve redundancy with the assumption that physiologically not all muscles are needed at the same time.
More sophisticated muscle recruitment strategies extensively studied in the literature can also be introduced in this reward function.
The last term prevents the agent from learning a so-called ``bang-bang'' solution~\cite{Spaces1980}, where a controller alternately switches between two extreme states, e.g.\  $\omega_\text{min}$ and $\omega_\text{max}$ herein. 
This term then ensures a sufficient exploration of the whole interval $[\omega_\text{min},\omega_\text{max}]$  during learning.
$P$ is inherent to the system being modeled, in our case the shoulder model and its forward simulation. 
$\gamma$ is a hyperparameter defining the discounting factor, set to be 0.99 herein.
To find an optimal policy, we comparatively study two following DRL strategies:

\emph{Deep Q-learning}
\cite{VanHasselt2016} is a common approach to find an optimal policy $\pi$ by maximizing the action value function, i.e.\ solving $Q^*(s_t) = \max_{\pi}Q_{\pi}(s_t,a_t) = r_t + \gamma \sum_{s_{t+1}\in S}P_{s_ts_{t+1}} \max_{a_{t+1}}Q^*(s_{t+1},a_{t+1})$, where $s_{t+1}$ and $a_{t+1}$ are respectively the next state and action, and $P_{s_ts_{t+1}}$ is the transition probability matrix for transitioning from state $s_t$ to $s_{t+1}$ for action alternatives.
$P$ can be populated by a so-called \emph{replay} buffer of past experiences, e.g.\ earlier simulations or games.
\emph{Deep Q-Learning} (DQL) approximates the Q-value function via a neural network (NN) and outputs discrete actions, typically converging relatively fast. 
As the action space of DQL, we quantized [-1,1]\% diffential activation range in 21 steps, i.e.\ $A_\text{DQL} = \{-1,-0.9,...,0.9,1\}\%$.
In other words, between two simulation steps any muscle activation cannot change more than 1\%. 

\emph{Policy Gradient Methods} work by estimating the policy gradient in order to utilize simple gradient-based optimizers such as Stochastic Gradient Descent for optimal policy decisions.
To prevent large policy changes based on small sets of data, Trust Region Policy Optimization (TRPO)~\cite{Schulman2015} proposes to regularize the policy update optimization by penalizing the KL divergence of policy change.
This helps update the policy in an incremental manner to avoid detrimental large policy changes, also utilizing any (simulated) training data optimally.
$Q_\pi$, $V_\pi$, and $\pi_\theta$ are estimated by different NNs, some parameters of which may be shared.
Policy gradient loss can be defined as $L^\mathrm{PG}(\theta) = \hat{\E}_{t} \left[\log \pi_{\theta}\left(a_t|s_t\right)D_t\right]$, where $D_t = Q_{\pi}(s_t,a_t) - V_{\pi}(s_t)$ is the \emph{advantage} function~\cite{Schulman2016}, which represents the added value (advantage) from taking the given action $a_t$ at state $s_t$.

Despite the constraint by TRPO, large policy updates may still be observed; therefore Proximal Policy Optimization (PPO)~\cite{Schulman2017} proposes to clip the gradient updates around 1 within a margin defined by an hyperparameter $\epsilon$ as follows:  
\begin{equation}
	L^\mathrm{CL}(\theta) = \hat{\E}_t \Big[\min \Big(\rho_t(\theta)D_t,\text{clip}(\rho_t(\theta),1-\epsilon,1+\epsilon)D_t\Big)\Big], \ \textrm{where} \ \rho_t(\theta) = \frac{\pi_{\theta}(a_t|s_t)}{\pi_{\theta_\text{old}}(a_t|s_t)}   \nonumber
\end{equation}
is the policy change ratio, with $\theta$ and $\theta_{\text{old}}$ being the policy network parameter vectors, respectively, before and after the intended network update.
The minimum makes sure that the change in policy ratio only effects the objective when it makes it worse.
NN parameters of the policy $\pi$ and the value function $V$ are shared, allowing to also introduce a value function error $L^\mathrm{VF}_t$. 
Additionally, to ensure policy exploration, an entropy bonus term is introduced~\cite{Schulman2017} as follows:
\begin{equation}
L(\theta) = \hat{\E} \left[L^\mathrm{CL}(\theta)- c_1L^\mathrm{VF}_t(\theta)+c_2S\left[\pi_{\theta}\right](s_t)\right],
\label{eq:L_ppo}
\end{equation}
where $c_1$ and $c_2$ are weights, $L^\mathrm{VF}_t = (V_\mathrm{new}(s_t)-V\mathrm{old}_t)^2$ is the change in value function before and after the NN update, and S denotes the entropy.
PPO can have continuous action spaces, so as its action set we used $A_\text{PPO} \in \left[-1,1\right]\%$ corresponding to same range for our DQL implementation.
In contrast to Q-learning with a replay buffer, PPO is an on-policy algorithm, i.e.\ it learns on-line via trial and error.

\vspace{1ex}\noindent\textbf{Implementation. }
We implemented\footnote{\url{https://github.com/CAiM-lab/PPO}} PPO~\cite{NEURIPS2019_9015} in Pytorch.
For DQL we used its OpenAI implementation~\cite{baselines}. 
For single-muscle control DQL and PPO were both implemented as simple networks of one hidden layer with 256 neurons.
For PPO with four muscles (PPO4), 3 hidden layers each with 250 neurons were used.
We used ReLu activations and the Adam optimizer.
For multibody biomechanical simulation, we used Artisynth~\cite{Lloyd}, a framework written in Java on CPU.
For training, the simulation runtime is the main computational bottleneck, with the network back-propagation taking negligible time.
For speed-up, we used a CPU cluster of 100 concurrent Artisynth simulations, each running a separate simulation episode and communicating with a DRL agent over a custom TCP interface based on~\cite{Abdi}. 
For simulation, an integration time-step of 100\,ms was chosen for a stability and performance tradeoff.
During training, at each simulation time step $t$ (DRL \emph{frame}), a simulation provides the respective RL agent with a state $s_t$ as in~(\ref{eq:reward}) including the simulated position $\phi(t)$ and velocity $\phi'(t)$ of the humerus.
The agent then calculated the respective reward $r(t)$ and, according to the current policy, executes an action, i.e.\ sends an update of muscles activations back to the simulation. 
This is repeated for a preset episode length (herein 10\,s), or until the simulation ``crashes'' prematurely due to numerical failure, which is recorded as a high negative reward.
Convergence was ascertained visually in the reward curves. 

\section{Experiments and Results}
Herein we demonstrate experiments showing a single-axis control of the shoulder.
The glenohumeral joint was thus modeled as a revolute joint allowing rotation only around the $z$ axis seen in Fig.\,\ref{MuskuloSkeletalModel}-center.
We conducted two sets of experiments:
In a preliminary experiment with only one muscle (ssp), we compared the presented RL algorithms for our problem setting.
A more sophisticated scenario with 4 muscles shows feasibility and scalability of the method to higher number of muscles. 
For training and testing, we used random trajectories.
Using 5$^{th}$-order polynomials~\cite{spong2020} as
$\hat{\phi}(t) = \sum_{i=0}^5 a_i\,t^i$\,, we generate 5\,s random sections. We set end-point velocity and acceleration constraints of zero, with end-point positions randomly sampled from [30,90]$^\circ$ during training, and from [20,100]$^\circ$ for testing.
Using a different and wider range for the latter was to test for generalizability.
By stacking such random sections while matching their end-point conditions, longer complex motions were constructed.
With this, for each training episode a 10\,s trajectory was generated on-the-fly, i.e.\ an episode being 100 frames given the integration time step of 0.1\,s.
Note that a trained RL agent can control an arbitrary trajectory length.
For testing, a set of 100 random trajectories of each 20\,s was generated once, and used to compare all presented RL agents; using root mean square error~(RMSE) and mean average error~(MAE) for tracking accuracy of desired trajectories. 

\vspace{1ex}\noindent\textbf{Control of Single Muscle Activation. }
With this experiment we aim to comparatively study DQL and PPO.
The exploration term in the reward~(\ref{eq:reward}) is irrelevant for DQL.
In order to have a fair comparison, we thus removed this term from the reward for this experiment.
Note that given a single muscle, the Lasso regularization of multiple activations in reward~(\ref{eq:reward}) also becomes unnecessary.
Accordingly, this experiment employs a straight-forward reward function as the absolute tracking error, i.e.\ $r(t+1) = -\big|\phi(t+1)- \hat{\phi}(t+1)\big|$. 

\begin{figure}[t]
\centering
\includegraphics[width=\textwidth]{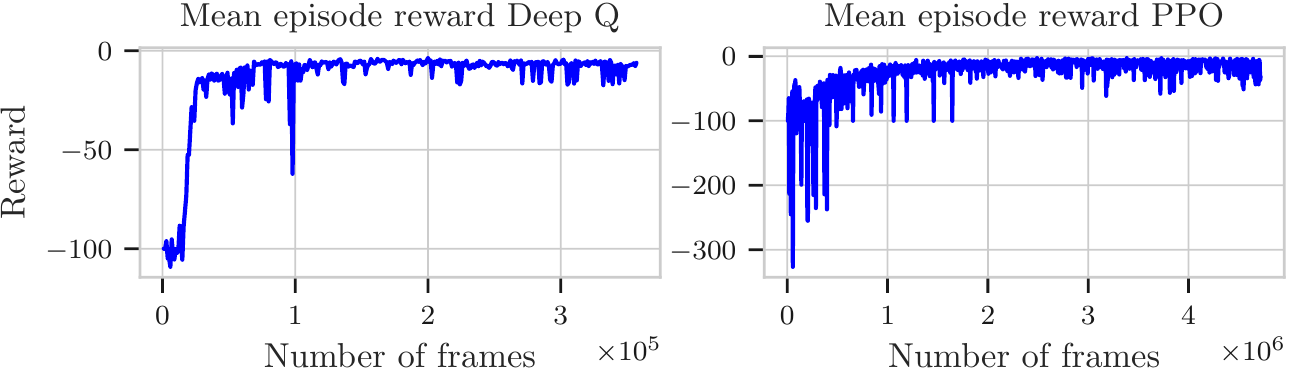}
\caption{Mean episode reward over last 10 episodes during training for DQL and PPO.}  \label{DeepQvsPPO}
\end{figure}

\begin{figure}[t]
\includegraphics[width=.805\textwidth]{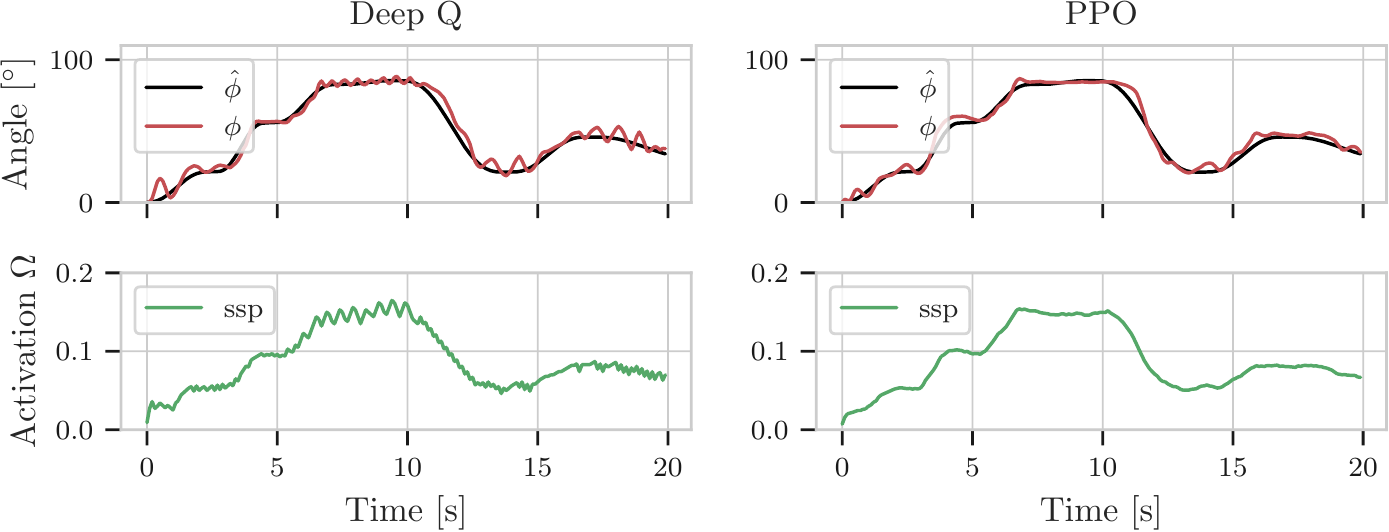}\hfill%
\includegraphics[width=.19\textwidth]{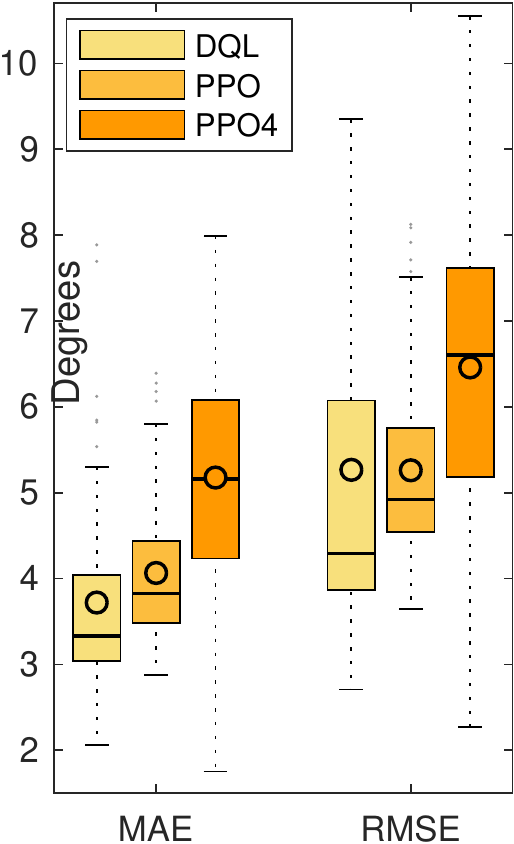}
\caption{A random target trajectory (black) and its tracking by RL using DQL (left) and PPO (center) with the muscle activation (bottom) and resulting angular upper arm motion (top).
(right) Distribution of tracking errors for 100 random trajectories.} \label{DeepQvsPPO2}
\end{figure}

Episode reward is defined as $\sum_{t=0}^{T}R(t)$ were T is the episode length.
Mean episode reward over last 10 episodes during training is depicted in Fig.~\ref{DeepQvsPPO} for DQL and PPO.
Both algorithms are observed to show a similar learning behaviour overall, although PPO requires approximately 10 times more samples than DQL to converge, due to $A_\text{PPO}$ being a continuous range. 
In Fig.~\ref{DeepQvsPPO2} both models are shown while controlling the ssp activation in the forward simulation for a sample trajectory. 
It is seen that the discrete action space of DQL results in a sawtooth-like pattern in activations $\Omega_\text{ssp}$, and hence a relatively oscillatory trajectory $\phi$.
Note that for small abduction angles the moment from humerus mass is minimal, and due to this lack of a counter-acting torsional load, the control becomes difficult, i.e.\ any slight changes in activations $\Omega_\text{ssp}$ may lead to large angular changes, visible in Fig.\ref{DeepQvsPPO2} for small abduction angles.

Over 100 trajectories, DQL has an MAE of 3.70$^\circ$ and RMSE of 5.78$^\circ$, while PPO has an MAE of 4.00$^\circ$ and RMSE of 5.36$^\circ$.
MAE and RMSE distributions of both methods over all tested trajectories can be seen in Fig.~\ref{DeepQvsPPO2}-right.

\begin{figure}[t]
\includegraphics[width=\textwidth]{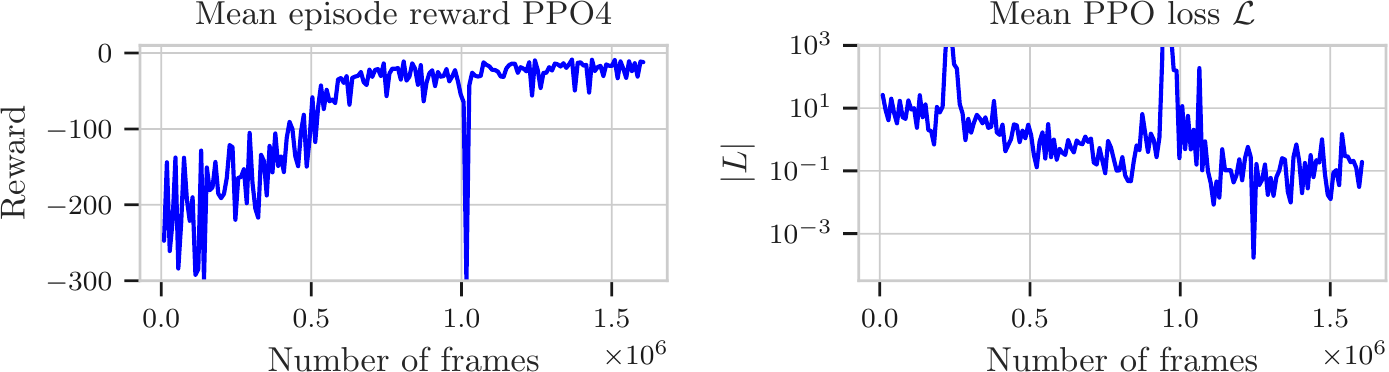}
\caption{Evolution of mean reward and loss of last 10 episodes during PPO4 training. } \label{DeepQvsPPO3}
\end{figure}

\begin{figure}[t]
\includegraphics[height=.36\linewidth]{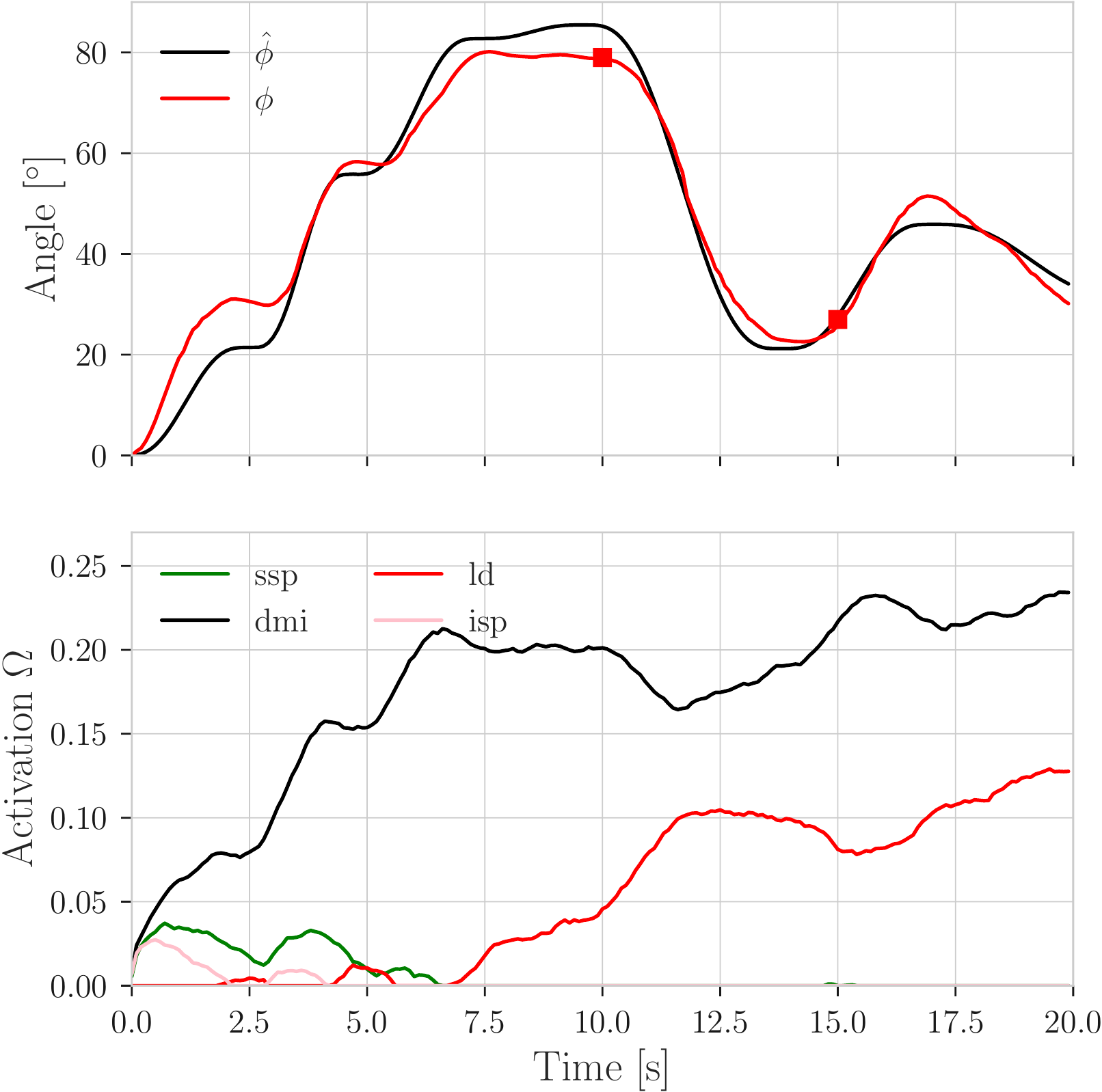}
\includegraphics[height=.36\linewidth,trim=20 0 0 0,clip]{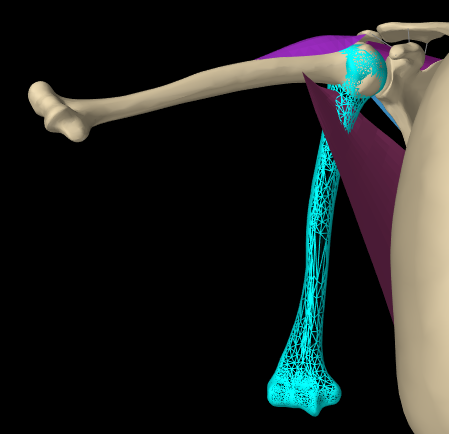}
\includegraphics[height=.36\linewidth,trim=0 3 30 0,clip]{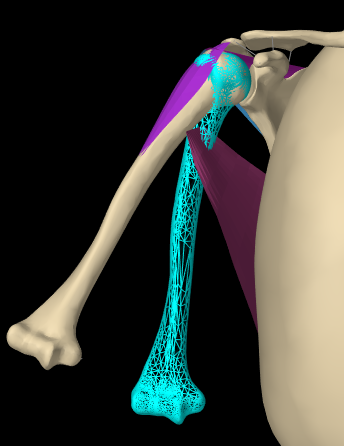}
\caption{Controlling four muscle activations for tracking the random abduction trajectory in Fig.\,\ref{DeepQvsPPO2}-left, along with the activation patterns (left).
Simulation frames at two time instances shown with red squares in the trajectory plot (center\&right).}
\label{realtimeTrajectory4}
\end{figure}

\vspace{1ex}\noindent\textbf{Muscle Control with Redundancy. }
In this scenario, all the four muscles relevant for abduction with redundancy are controlled at the same time.
Given similar action space quantization of 21 steps, DQL for 4 muscles would require a $4^{21}$ dimensional discrete action space $A_{\text{DQL}}$, which is computationally unfeasible.
Indeed, this is a major drawback of DQL preventing its extension to high dimensional input spaces.
In contrast, a continuous action space for PPO is easily defined for the four muscles.
Given the simulation with 4-muscles, a PPO agent (PPO4) was trained for 1.6\,M frames, taking a total of 1 hour including overheads for communication and Artisynth resets after crashes.
Mean episode reward and loss~(\ref{eq:L_ppo}) averaged over last 10 episodes are plotted in Fig.~\ref{DeepQvsPPO3}.
Note that high gradients in policy updates due, e.g., to crashes, is a challenge 
Despite the 4 times higher action space dimension, a feasible learning curve is observed.
Large negative spikes in reward, e.g.\ near 1\,M frames, correspond to simulation crashes, e.g.\ due to infeasible activations generated by the DRL agent.
Despite the large policy gradients these spikes cause in~(\ref{eq:L_ppo}), PPO is able to successfully recover thanks to its gradient clipping. 
Using the trained PPO4 agent for controlling four muscles, Fig.\,\ref{realtimeTrajectory4} shows the humerus tracking for the same earlier random trajectory in Fig.\,\ref{DeepQvsPPO2}, with the PPO-generated muscle activations.
It is observed that ssp and isp help to initiate the abduction motion -- a well-known behaviour~\cite{Reed2013}. 
Beyond initial abduction, their activation however diminishes with the rest of the motion mainly carried out by dmi and ld, which have stronger moment arms. 
PPO control of 4-muscles during 100 randomly-generated trajectories results in an MAE of 5.15$^\circ$ and RMSE of 6.64$^\circ$, with their distributions shown in Fig.~\ref{DeepQvsPPO2}-right.
The slightly higher tracking error compared to single-muscle case is likely due to higher network capacity, training episodes, and thus time for convergence required for a higher dimensional action space.

We further tested an in-vivo trajectory from the public motion-tracking dataset of~\cite{Bolsterlee2014}.
We used a \emph{combing} motion of 17.5\,s, involving lifting up the arm twice and combing while up. 
Using the angle between the humerus and vertical axis, the 3D tracked motion was converted to an abduction angle as our tracking target (varying between 20 and 100 degrees). 
Applying our earlier-trained PPO4 agent on this shows good tracking visually, with an RMSE and MAE of 7.67$^\circ$ and 6.57$^\circ$.
These results being comparable with the earlier ones show that our method generalizes well to in-vivo trajectories even with synthetic training.

\section{Conclusions}
We have studied two DRL approaches demonstrating the successful application for single-axis control of a functional biomechanical model of the human shoulder.
PPO was implemented in a way that allows for multiple environments to run simultaneously using network based sockets. 
This is indispensable for scalability to higher dimensions within reasonable computational time-frames. 
Inference of our NN-based DRL agents are near real-time, enabling fast control of complex functional simulations. 
Any constraints that make tracking suboptimal or simulation infeasible are implicitly learned with DRL, as a remedy to occasional simulation crashes occurring with conventional analytical controllers.

A main bottleneck for scalability to sophisticated models is the limitation with action spaces.
In contrast to the discrete action space of DQL exploding exponentially with the curse of dimensionality, it is shown herein that the continuous action space and corresponding policy optimization of PPO enables its extension to multiple muscles with redundancy.
Given the generalizable form of the learning scheme and the reward function with the proposed approach, extensions to more muscles and additional degrees-of-freedom is straightforward.
This opens up the potential for full control of the shoulder and other musculoskeletal structures. 
This also enables neuroplasticity studies after corrective surgeries such as muscle transfers:
After major orthopedic interventions, the patients may not easily adjust to postop configurations, therewith recovery expectancy and rehabilitation time-frames varying widely. 
Networks trained on preop settings and tested on simulated postop scenarios could provide insight into operative choices, e.g.\ for faster rehabilitation and improved outcomes.

\bibliographystyle{splncs04}
\bibliography{DRLshoulder}

\end{document}